\documentclass[aip,amsmath,amssymb,12pt
]{revtex4-1}

\usepackage{soul}   % for \st

\makeindex

\makeatletter
\newcommand\chaptercontents{
{\global
\@topnum\z@ 
\@afterindentfalse 
\if@twocolumn
\@restonecoltrue
\onecolumn 
\else 
\@restonecolfalse 
\fi 
\vspace*{10pt}
\noindent 
{\small\bf Contents}\par 
\vskip1em 
\nobreak} 
{\small
\@starttoc{toc}%
}\if@restonecol
\twocolumn
\fi}
\renewcommand*\l@section[2]{% 
\ifnum \c@tocdepth >\z@
\addpenalty\@secpenalty 
\setlength\@tempdima{2.5em}% 
\begingroup 
\parindent \z@ 
\rightskip
\@pnumwidth \parfillskip -\@pnumwidth 
\leavevmode 
\advance\leftskip\@tempdima 
\hskip -\leftskip 
#1\nobreak\leaderfill\nobreak
\hb@xt@\@pnumwidth{\hss #2}\par 
\endgroup 
\fi}
\renewcommand*\l@section{\@dottedtocline{1}{0.1em}{1.3em}}       %for 3
\renewcommand*\l@subsection{\@dottedtocline{2}{1.5em}{2em}}      %for 3.1
\renewcommand*\l@subsubsection{\@dottedtocline{3}{3.5em}{2.6em}} %for 3.3.1

\makeatother

\usepackage{amsfonts}
\usepackage{amsmath}
\usepackage{amssymb}
\usepackage{amsthm}
\usepackage{appendix}
\usepackage{chemformula}
\usepackage{datetime}
\usepackage{dcolumn} % Align table columns on decimal point
\usepackage{enumerate}
\usepackage{enumitem}
\usepackage[T1]{fontenc}
\usepackage{epigraph}
\usepackage[T1]{fontenc}
\usepackage{graphicx} % Include figure files
\usepackage[utf8]{inputenc}
\usepackage{hyperref}
\hypersetup{
    colorlinks=true,
    linkcolor=blue,
    filecolor=magenta,      
    urlcolor=cyan,
}
\usepackage{mathptmx}
\usepackage{tensor}
\usepackage{xcolor}
\usepackage{yfonts}

\newcommand{\R}{\mathbb{R}}  % real numbers
\newcommand{\N}{\mathbb{N}}  % natural numbers

\newcommand{\bfr}{{\bf r}}

\begin{document}

\title{
%Putting Maximum Probability Domains in the Context of Other Methods Describing Chemical Bonding
What is the number of electrons in a spatial domain?}

\author{Anthony Scemama} 
\affiliation{Laboratoire de Chimie et Physique Quantiques (UMR
  5626), Université de Toulouse, CNRS, UPS, France \\
  scemama@irsamc.ups-tlse.fr}
\author{Andreas Savin}
\affiliation{CNRS and Sorbonne Université, Laboratoire de Chimie Théorique (LCT),\\
  F-75005 Paris, France\\
  andreas.savin@lct.jussieu.fr}

\markboth{A. Scemama, A. Savin}{Number of electrons}

\begin{abstract}
We like to attribute a number of electrons to spatial domains (atoms, bonds, \dots).
However, as a rule, the number of electrons in a spatial domain is not a sharp number. 
We thus study probabilities for having any number of electrons (between 0 and the total number of electrons in the system) in a given spatial domain.
We show that by choosing a domain that maximizes a chosen probability (or is close to it), one obtains higher probabilities for chemically relevant regions.

The probability to have a given electronic arrangement, -- for example,
by attributing a number of electrons to an atomic shell  -- can be low.
It remains so even in the "best" case, i.e, if the spatial domain is chosen to maximize the chosen probability.
In other words, the number of electrons in a spatial region significantly fluctuates.

The freedom of choosing the number of electrons we are interested in shows that a "chemical" question is not always well-posed.
We show it using as an example the \ch{KrF2} molecule.
\end{abstract}


\maketitle

\tableofcontents
\newpage

\centerline{\today \hspace{2pt} at \currenttime}

\section{Introduction}

\subsection{Chemical introduction to the subject}
When speaking about chemical bonding, it is useful to make a distinction between effects that are discussed at the collective (molecular, crystalline) level, and those associated to a fragment.
For example, the energy lowering obtained when atoms form a molecule is providing information that we qualify here as collective.
However, there are many quantities that are obtained when looking at pairs of atoms forming a molecule, for example: 
\begin{itemize}
 \item lines drawn connecting atoms, e.g, C-H, since the 19th century.
 \item the energy attributed to a pair of them (e.g., that of the CC or CH bonds in saturated hydrocarbons), 
 \item nearly invariant distances between types of atoms, e.g., the length of the CH bond,
 \item patterns to understand spectra, e.g., attributed to the CH stretching frequency,
 \item \dots
\end{itemize}

In this paper we are interested in describing grouping of electrons in some spatial domain, $\Omega$. 
We use quantum mechanical calculations, and start with the Schrödinger equation.
The Hamiltonian gives a natural partitioning, and it is reasonable to use it (see, e.g.~\cite{Hel-33,Rue-62}), as well as an energy partitioning resulting from it.
In many cases the physical origin of the formation of groups is the Pauli principle. 
This directs us toward analyzing the wave function.
Due to the complicated structure of the wave function, its reduction to three-dimensional objects is desired.
It is worth mentioning in this direction the work of Artmann~\cite{Art-46} and that of Daudel~\cite{Dau-53}.
Using the electron density, $\rho$, as proposed by Bader for the Quantum Theory of Atoms in Molecules (QTAIM)~\cite{Bad-90} had, and still has a great success.
The Pauli principle is hidden in the density.
It is made more explicit~\cite{SavJepFlaAndPreSch-92} in the Electron Localization Function (ELF) of Becke and Edgecombe~\cite{BecEdg-90}.
The Maximum Probability Domains, MPDs~\cite{Sav-02}  (or their simplified variants), of interest in this paper, originate from Daudel's idea of partitioning 3D space (into so-called "loges"), using the wave function squared. 
However, instead of making a partition of the whole molecular (or crystalline) space, with MPDs one concentrates on specific spatial regions, thus reducing the computational effort, and avoiding the propagation of errors produced in a region different from that of interest.

\subsection{Quantum mechanical introduction to the subject}

We speak about having two electrons in a bond, eight electrons in the valence shell of Ne, atomic charges, and so on.
% As we consider regions not covering the whole space, the number of electrons in them is not sharply defined.
The operator that gives the number of electrons in a spatial domain, $\Omega$, is
\begin{equation}
\label{eq:n-hat}
    \hat{N}(\Omega) = \int_\Omega \hat{\rho}(\bfr_i - \bfr) \, d \bfr
\end{equation}
where 
\begin{equation}
    \hat{\rho}(\bfr) = \sum_{i=1}^N  \delta(\bfr_i-\bfr)
\end{equation}
is the density {\em operator}, $N$ the total number of electrons in the system, $\delta$ is Dirac's $\delta$ function, $\bfr_i$ are the positions of the electrons, and $\bfr$ refers to an arbitrary position in the three-dimensional space.
The eigenfunctions of the Hamiltonian operator are not, in general, eigenfunctions of $\hat{N}_\Omega$. 
\footnote{The operators do not commute for arbitrary $\Omega$, while they trivially commute for $\Omega$ being the whole space, as in this case $\hat{N}$ becomes $N$, the number of electrons in the system.} 
As a result, we cannot specify a given number of electrons in $\Omega$.
However, we can specify a {\em probability} to have a given number of electrons in $\Omega$.

In this paper we choose a number of electrons, $n$.
It is provided by chemical intuition, e.g., of having eight electrons in the valence shell of the Ne atom.
We are interested in the spatial region that maximizes the probability of having that chosen number of electrons in it.
This is a {\em Maximum Probability Domain} (MPD, see appendix~\ref{app:mpd} for details). 
Note that the probabilities of finding an arbitrary number of particles can be obtained for any spatial region, $\Omega$, for example in the basins of the electron density as provided by  QTAIM~\cite{Bad-90}, or those of the electron localization function, ELF.~\cite{ChaFueSav-03}
Note that an error produced by an approximation to a MPD produces only second order errors in the probabilities, because the probability is maximal for an MPD.

As with localized orbitals one may consider electron pairs, and obtain spatial regions that can be associated to one or more nuclei (lone pairs, two-center bonds, three-center-bonds, dots).
However, the number of electrons considered for an MPD, $n$,  can be adapted to the question of interest.
For example, one may want to search for a given ion in a crystal and choose $n$ equal to the number of electrons in that ion.~\cite{CauSav-ion-11}
Furthermore, one may consider spatially disconnected regions, for example when considering spin couplings of electrons on different centers.

The definition of the probabilities and the MPDs use the wave functions squared.
Thus, there is no restriction to ground states.
The same definitions can be applied to time-dependent processes.~\cite{Sav-18}

One can consider probabilities for multiple domains, e.g., establish connections to resonant structures.~\cite{PenFraBla-b-07,PenFraBla-c-07}
It is possible to define a joint probability (of having $n_A$ electrons in $\Omega_A$, and $n_B$ electrons in $\Omega_B$), or a conditional probability (of having $n_A$ electrons in $\Omega_A$ given that there are $n_B$ electrons in $\Omega_B$).~\cite{Dau-53, GalCarLodCanSav-05, PenFraBla-a-07, PenFra-19, SceSav-22}

\section{How numbers are obtained}
\subsection{Probabilities}
\subsubsection{Choosing the relevant quantities to be obtained}

Traditionally, one looks at the population of a spatial region.
We can see from Eq.~\ref{eq:n-hat} that the expectation value of the number of particles in the domain $\Omega$ is just its population,
\begin{equation}
    \label{eq:pop-mean}
    \langle \Psi | \hat{N}(\Omega) | \Psi \rangle = \int_\Omega \rho(\bfr) d\bfr
\end{equation}
$\Psi$ is the wave function of the system, $\rho$ its one-particle density, and $N$ the number of electrons in the system.
Being a mean value, we can express it in terms of probabilities.
\begin{equation}
    \label{eq:pop-p}
    \mu(\Omega) = \langle \Psi | \hat{N}(\Omega) | \Psi \rangle = \sum_{n=0}^N n \, P(n,\Omega)
\end{equation}
$P(n,\Omega)$ is the probability to have $n$ electrons in $\Omega$.
Its expression is given in appendix~\ref{app:mpd}.

In the same way, we can express the variance,
\begin{equation}
 \label{eq:sigma}
 \sigma^2(\Omega) = \sum_{n=0}^N (n-\mu)^2 P(n,\Omega)
\end{equation}

\begin{figure}[htb]
 \begin{center}
   \includegraphics[width=0.95\textwidth]{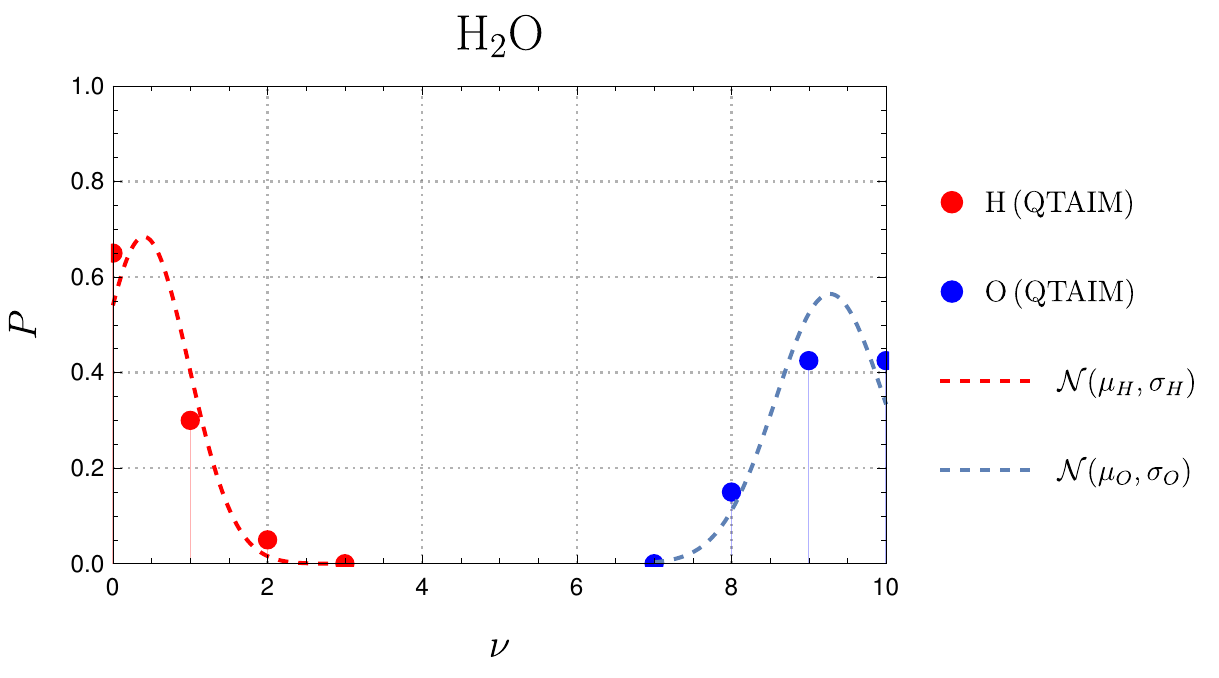}
 \end{center}
 \caption{Probability distribution for the atomic (QTAIM) basins of O and H in the water molecule (dots) and the normal distributions (dashed lines) with the same $\mu$ and $\sigma$.~\cite{ChaFueSav-03}
 The results for H are in red, for O in blue.}
 \label{fig:p-h2o-aim-be}
\end{figure}

It is tempting to indicate just $\mu$ and $\sigma$.
If the distribution of probabilities were normal, $\mu$ and $\sigma$ would be sufficient to recover all information about the distribution of probabilities.
However, the distribution is not normal.
As $n$ is always an integer, we do not have a continuous probability distribution, so it cannot be a normal distribution.

To avoid this argument, we can argue that we use only a Gaussian function of a real $n$, but read it only at integer values of $n$ to obtain the values of the probabilities.
In many cases, this is expected to work well (cf. Ref.~\cite{PfiBohFul-85}).
However, there is also another aspect to consider with normal distributions: a normal probability distribution function is non-zero for arguments that extend to negative values and to values larger than $N$.
This is physically impossible.
We should restrict the reading on the Gaussian curve only to values $n=0,1,\dots,N$.
Let us take as a numerical example, where $\Omega$ is an atomic basin (QTAIM) in the water molecule, Fig~\ref{fig:p-h2o-aim-be}.~\cite{ChaFueSav-03}
Choosing just points for integer values on a normal distribution gives the absurd interpretation that there is a significant probability ($\approx 0.2$) to have -1 electron in the H atom basin.
Furthermore, there is a similar probability to have 11 electrons in the O atom basin (10 being present in the water molecule).

Fig.~\ref{fig:p-h2o-aim-be} can induce us to believe that the probability to find $n$ electrons in $\Omega$ could be read (up to a precision of about 0.1) at admissible values of a normal probability distribution function with the same mean and variance as provided by the physical probability distribution .
However, let us consider now the dissociation of the H$_2$ molecule.
The covalent (ground state) dissociation produces for $\Omega$ being the half space containing a H atom, $P(n=1,\Omega)=1$, yielding $\mu=1,\sigma=0$. 
In this case, indicating $\mu$ and $\sigma$ is sufficient.
A different situation arises when we consider a state that dissociates into the ionic form, H$^+$~\dots~H$^- \leftrightarrow$~H$^-$~\dots~H$^+$. 
We get $P(n=0,\Omega)=P(n=2,\Omega)=1/2$, $P(n=1)=0$. 
The mean (the population) is the same as for the covalent case, $\mu=1$, while the variance is different, $\sigma^2=1/2$.
A Gaussian form with the mean at $\mu=1$ yields a maximum (0.56) at $n=1$, where $P$ is 0, and too low estimates at $n=0,2$, namely, 0.21 instead of 0.5.

In statistics, more information from probability distributions is summarized by introducing 
higher order (standardized) moments, e.g the third power of $(n-\mu)$ (skewness) or the fourth power (kurtosis).
However, if we look at the data, we see that the number of cases where the probabilities $P(n,\Omega)$ significantly differs from zero is small, and already contains all the relevant information. 
Thus, we may use directly the significant $P(n,\Omega)$ instead of using statistical summaries.
If needed, the latter can be easily obtained once the $P(n,\Omega)$ are known, as, for example, in Eqs.~\eqref{eq:pop-p} and \eqref{eq:sigma}.

\subsubsection{Computing $P(n,\Omega)$}

For Slater determinants (as obtained from Hartree-Fock or Kohn-Sham calculations), $P(n,\Omega)$ can be computed from the overlap integrals
\begin{equation}
    S_{ij}(\Omega) = \int_\Omega \phi_i(\bfr) \phi_j(\bfr) d\bfr 
\end{equation}
$\phi_i, \phi_j$ are the orbitals present in the Slater determinant.~\cite{Sav-02,CanKerLodSav-04} 
Note that here the integration is not performed over $\R^3$, but over $\Omega \subset \R^3$.
For large systems, it is convenient to use localized orbitals in order to neglect $S_{ij}$ between distant orbitals (that do not overlap in $\Omega$).

Multi-determinant wave functions can be also used.~\cite{FraPenBla-07, FraPenBla-08,PenFraBla-a-07}
Quantum Monte Carlo calculations are very flexible in the choice of wave functions and are convenient for estimating $P(n,\Omega)$.~\cite{SceCafSav-07} Moreover, computing the probabilities with samples drawn from a Monte Carlo sampling of the squared wave function is particularly simple: 
one simply counts the number of electrons present in $\Omega$ for each configuration generated during the calculations.
The ratio between the number of configurations presenting $n$ electrons in $\Omega$ and the total number of configurations is an estimator of $P(n,\Omega)$.
As a rule, the number of configurations needed to obtain a reasonable probability is much lower than that for obtaining a reasonable energy simply because the number of digits needed is much lower for probabilities.

\subsubsection{Sensitivity to the choice of the wave function}
All the interpretative methods raise the question whether the refinement of the method (such that the change of the basis set) significantly changes the conclusions. 
The simplest wave function capturing the physics should be sufficient.
In the case we discuss, the Pauli principle is already described by a single determinant wave function, so methods like Hartree-Fock or the Kohn-Sham method should be sufficient in most cases.
For example, it is known that the density can be reasonably obtained with relatively low level methods (defining atomic basins in QTAIM). 
As a counter-example, consider $|\nabla \rho|/\rho$.
It is a good detector of the shell structure in atoms, but its topology is sensitive to the (Gaussian) basis set used.~\cite{KohSavPre-91}

In many cases, using a correlated wave function does not change significantly the probabilities.
MPDs show often little sensitivity to the wave function used -- as long as the Pauli principle is the underlying cause of the property studied.
Nevertheless, there are cases (of near-degeneracy) when correlation effects are felt, and multi-determinant wave functions that describe correctly the situation should be better used.
For example, at dissociation, the \ch{H2} molecule yields at Hartree-Fock level $P(n=1,\Omega)=1/2$, where $\Omega$ is the half-space defined by a plane perpendicular to the H-H axis, at the midpoint between the nuclei.
Correlation effects can be seen also at equilibrium distance.
For example, let us consider the \ch{F2} molecule.
We divide again the space between the two atoms by choosing a plane perpendicular to the molecular axis, at equal distance from the two nuclei.
For the correlated wave function, we obtain for $\Omega$ corresponding to the half-space
$P(n=9)=0.61$, $P(n=8)=P(n=10)=0.19$,
while from the Hartree-Fock we find a higher importance of the ionic functions,
$P(n=9)=0.47$, $P(n=8)=P(n=10)=0.25$.

Details concerning the wave functions used below can be found in appendix~\ref{app:comput}.
Some of the wave functions used are at Hartree-Fock level, some can be considered quite accurate.
We do not expect qualitative changes in our discussion by further improvement of the wave function.

\subsection{Spatial domains}
\subsubsection{MPDs are not basins}
When using $\rho$, or ELF one has functions defined in 3D.
The spatial regions are obtained by constructing basins.~\cite{Bad-90,SilSav-94}
To obtain MPDs one directly constructs the spatial regions, by choosing to maximize the probability to have $n$ electrons in it.

\subsubsection{Known spatial regions}
There are limiting cases where MPDs are known.
\begin{itemize}
    \item The probability to have all $N$ electrons in $\Omega$ is maximal when $\Omega$ is the whole space.
    We are sure that all electrons are in it, $P=1$. 
    \item When the volume of $\Omega$ is vanishing, we know that there are no electrons in it. 
    In this case, $P \rightarrow 1$.
    \item Spatial regions can be equivalent by symmetry. 
    If we have determined the MPD for one of the elements that are equivalent by symmetry, we can obtain all the others by performing symmetry operations.
    \item If we have determined the MPD for a given $n$, $\Omega_n$, the MPD for $N-n$ electrons is the remaining space;
    $P(n,\Omega_n)=P(N-n,\R^3\backslash\Omega_n)$.
\end{itemize}

\subsubsection{Shape optimization}
Algorithms to deform $\Omega$ to maximize the probability exist (see, e.g., ~\onlinecite{CanKerLodSav-04, BraDalDapFre-20}).
Such (shape-optimization) algorithms even allow having spatial regions that are not connected.
One starts with a given spatial domain, $\Omega$, and computes $P(n,\Omega)$.
$\Omega$ is slightly deformed to increase the $P(n,\Omega)$, until the latter is maximized.
Unfortunately, there are some drawbacks.
\begin{itemize}
    \item The programs to compute the MPDs are not widely distributed.
    \item The algorithms are computationally demanding, in spite of the fact that $\Omega$ has to be considered in a restricted region of space.
    \item For algorithms based on quantum Monte Carlo sampling, there are difficulties when the number of  configurations is low at the separation surface or $\Omega$, introducing some uncertainty.
\end{itemize}

One may treat the last two problems by using smooth boundaries instead of having sharp boundaries.
The smoothing functions can depend on parameters that could be optimized directly.
One should keep in mind that smoothing the borders can lower the
probabilities.~\cite{SceSav-22} At first, this may seem counter-intuitive.
To understand it, one can imagine smoothing the boundaries of the MPD for $n$ electrons, $\Omega_n$, as mixing to some degree spatial regions for which $P(n,\Omega)$ is lower.

\subsubsection{Partial optimization of the domain}

Recall that -- when we are close to the maximizing domain, $\Omega_n$ -- the errors in $P(n,\Omega)$ are only of second order in the change between $\Omega$ and $\Omega_n$.
Thus, instead of smoothing the boundaries, we can stop before reaching the full optimization of $\Omega$.~\cite{SceSav-22}

One way to do it is to define specific shapes, and optimize a reduced number of parameters.
Let us give some examples of such incomplete optimization.
\begin{itemize}
    \item In a molecule, the atomic core is not identical to the spherical one  obtained for the isolated atom.
    However, we can assume that it can transferred from the atom.
    In all cases treated so far, the difference observed is at most in the second decimal of the probabilities.
    \item One can define points in space that are used to define "centers" around which the MPDs are constructed as Voronoi cells. 
    The positions of the centers can be varied, in order to maximize the probabilities.
    More flexibility may be gained by modifying the definition of distances, e.g., by introducing weights.
    \item Often, one can use a good guess for MPDs, e.g., ELF basins.
\end{itemize}

\begin{figure}
    \centering
    \includegraphics{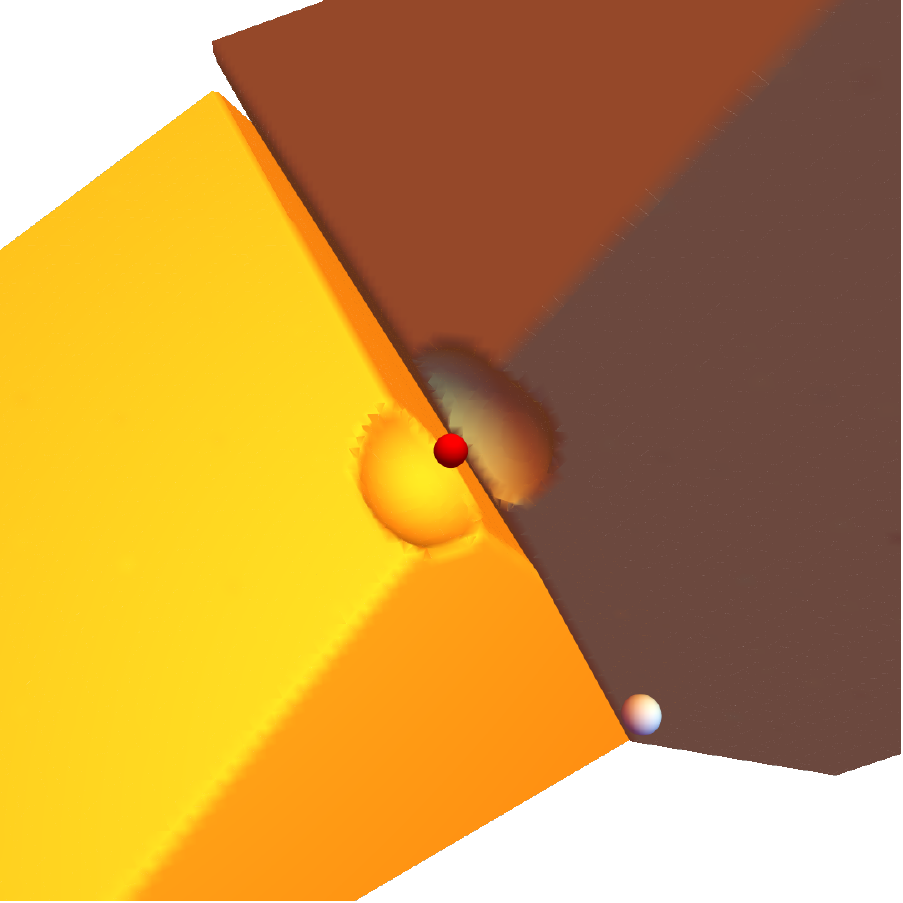}
    \caption{Spatial domains that maximize the probability for a domain associated to the OH bond electron pair (orange), or the O lone pair (brown), by excluding a region associated to the O core, and by defining Voronoi cells with moving centers. The positions of the nuclei are indicated by small spheres (red: O, white: one of the H atoms, the other being hidden behind the orange surface).}
    \label{fig:voronoi}
\end{figure}

Let us take as an example the construction of domains in the \ch{H2O} molecule.
We first determine the core domain, by maximizing the probability to have 2 electrons within a sphere around the O nucleus.
For a radius of 0.36 bohr, we obtain the maximal probability 0.73.
We choose four points; two are in the plane defined by the plane of the nuclei, two are in the plane perpendicular to the previous one.
For example, we may start with a tetrahedron having two vertices on the H nuclei, and the O nucleus in the center.
The centers define Voronoi cells.
We further exclude the core domain, and compute the probability of having 2 electrons in one of the regions containing a H atom.
We now change the positions of the points, respecting the symmetry of the molecule, to maximize the probability.
The maximum is reached with $P=0.45$. 
We can repeat this procedure for maximizing the probability of having two electrons in the region that corresponds to one of the lone pairs (one of the centers in the plane perpendicular to the HOH plane).
The maximum is reached with $P=0.41$. 
The probabilities obtained this way do not differ by more than 0.01 from those obtained after full optimization.
The domains obtained are shown in Fig.~\ref{fig:voronoi}.

It is also interesting to consider several spatial regions, e.g., for analyzing statistical correlations between them~\cite{GalCarLodCanSav-05, PenFraBla-a-07, PenFra-19} or electrons distributed over disconnected spatial regions.~\cite{Sav-21}
A problem that appears when considering joint probabilities is that they are lower than that of the individual ones.
Recall that for independent events, the probability of the joint event is the
product of probabilities; as the probabilities are between 0 and 1, their
product is lower than each of the individual probabilities.  In this context,
it is more reasonable to consider conditional probabilities, e.g., the
probability to have $n_A$ electrons in $\Omega_A$ given that there are $n_B$
electrons in $\Omega_B$~\cite{Dau-53, GalCarLodCanSav-05, PenFra-19,SceSav-22},
\begin{equation}
    P_{A|B}= \frac{P_{A \cap B}}{P_B}.
\end{equation}

\subsubsection{Describing the spatial domains}
Of course, the spatial extension of the MPD can be graphically shown, and this is consistent with the pictorial attitude existing in chemistry.
Some numbers can also be used to describe them when the parametrization of the domain is simple.
For example, the core region of the Ne atom can be represented by a sphere of radius $r$ maximizing the probability of having two electrons, and the valence region is the complement. The probability is maximal for $r \approx 0.27$~bohr. Similarly, for diatomic molecules a single number is sufficient to describe the position of the plane perpendicular to the molecular axis, which 
defines the boundary between the two atomic domains.

\subsubsection{Multiplicity of MPDs}
\label{multiplicity}
For a given molecule, the MPDs are defined by indicating a number $n$ of electrons in them, and are obtained by optimization of the spatial domain.
The latter process can lead to several solutions (several local maxima may exist).
In this respect, the MPDs behave in a way similar to localized orbitals: equivalent solutions exist.

There are trivial cases.
For example, if we search in the H$_2$O molecule for a MPD for $n=2$ electrons, we can find a domain corresponding to the core, or to any of the OH bonds, or any of the two lone pairs.
Note that some of these solutions have chemically different significance, e.g., core \textit{vs} lone pairs. 
Other solutions may be equivalent by symmetry e.g., the MPDs corresponding to the two OH bonds.

Notice the analogy to localized orbitals.
These also can lower the symmetry, and equivalent solution exist.
A simple example is that of the $\pi$ orbitals in benzene, where 
 one has three localized orbitals for a six-fold axis.
 
Sometimes one needs a moment of reflection to discover this effect.
For example, in trans-HSiSiH one has three electron pairs connected to the two Si atoms.
However, the system is invariant under the inversion operation ($\bfr \rightarrow -\bfr$) that is lost once a set three MPDs are found; the symmetry operation produces another equivalent set.~\cite{SceCafSav-07}

In general, we expect a displacement of the MPD to lower the probability associated to it.
For example, transforming the MPD into another $\Omega$ by inversion through the position of the C nucleus lowers the probability from 0.55 to 0.36.
However, an infinite set of equivalent solutions can be produced by symmetry.
This also presents an analogy to localized molecular orbitals.~\cite{Eng-71,EngSalRue-71}
For example, in the HCCH molecule, we can find a banana bond between the two C atoms, but cylindrical symmetry dictates that any rotation around the molecular axis yields an equivalent solution.
The same type of situation appears in atoms, e.g., the Ne atom. 
When searching for a pair of electrons we will find a domain avoiding the core region and resembling to an $sp^3$ hybrid, pointing into an arbitrary direction.
However, any rotation with the center on the Ne nucleus produces an equivalent MPD.
In the uniform electron gas, any translation produces equivalent MPDs.
One expects that in metals  deformations of the MPD have little effect on the probability.
A study of the Kronig-Penney model gives a hint in this direction.~\cite{Sav-21}

Methods like ELF give in such cases solutions that average out the effect of different solutions: one gets a single bonding region for the CC bond in acetylene, a valence shell for the Ne atom, a constant value through the uniform electron gas.
However, in the case of HSiSiH discussed above, this "averaging out" produces four basins. 
This is disturbing, because there are only three bonds.~\cite{SceCafSav-07}

In the case of MPDs, one can consider larger groups, e.g., $n=6$ electrons for the triple CC bond in HCCH, or $n=8$ electrons for the valence shell of Ne. 
%If one wishes, one may consider reducing the symmetry (e.g., by perturbing the system with a reactant).

\section{What the numbers tell us: comforting and disturbing results}
\subsection{Comforting results}
\subsubsection{Conceptual advantage}
The MPDs are simple to explain, applicable to simple or complicated wave functions.
As the number of electrons in a spatial domain is user-defined, one is not compelled to study a given object.
For example, one can use MPDs to find electron pairs, bonding regions in diamond, as with ELF~\cite{CauSav-11}, or to find ions in crystals, as with QTAIM~\cite{CauSav-ion-11}.
In many instances it gives results that are consistent with those obtained with other methods, such as QTAIM or ELF.
This can be seen, for example when looking at crystals in rock salt structure (when recognizing ions)~\cite{CauSav-ion-11}, or at crystals with diamond structure (for covalent bonds)~\cite{CauSav-11}. 
This is very encouraging, taking into account the wide success of QTAIM and ELF.

\subsubsection{Producing reasonable numbers}
We are used to look at populations (as defined in Eqs~\ref{eq:pop-mean},\ref{eq:pop-p}).
While numbers obtained with different approaches can slightly differ, there is some consensus about what we should expect from some populations: that of "standard" bonds should be close to 2, that of atomic shells, etc.

The population cannot be used to define a spatial region;
one cannot define atomic shells by requesting that the number of electrons integrate to a specified number.
For example, we can find in Be an infinity of spherical shells, between $r_{min} \in(0,r_{core})$ and $r_{max} \in (r_{core}, \infty)$, where $r_{core} \approx 1$~bohr,
just requesting that the integral of the density between $r_{min}$ and $r_{max}$ equals two.

One may ask whether all methods give equivalent results.
For example, it would not be worth computing the MPD if ELF and the Laplacian of the density would give the same result.
Very often, the MPDs are close to other spatial regions, e.g., the ELF basins when searching for regions characterizing electron pairs.
However, it is known that the shell structure of atoms is not always correctly reproduced by the Laplacian of the density;
for example, the last shell of the Zn atom is merged with the penultimate shell.~\cite{Bad-90}
ELF separates them, but the population of the valence shell is 2.2 instead of 2.~\cite{KohSav-96}
With MPDs, the difference between the expected population of the valence shell and that obtained is at the second decimal, 1.96 with the Hartree-Fock wave function. 
(This is also roughly the accuracy we expect for the data discussed in this article.)
The last shell is also separated in atoms like Nb, or Mb yielding in these cases a population of 1.0.~\cite{Sav-02}.

In molecules like \ch{CH4} or \ch{H2O}, MPDs define regions of space that are conventionally attributed to the bonding or the lone pairs.
The populations are very close to the expected number, 2. 
Even when the electrons are "crowded", like in the \ch{N2} molecule, the population is not too far from 2 (it is 2.2).

There can be also qualitative differences, between, say, ELF and MPD results, in particular when several alternative classical bonding situations exist~\cite{SceCafSav-07}.
Differences may appear because ELF is producing spatial domains that respect symmetry, e.g., the spherical shells in atoms, while there may be several ways MPDs can be produced.
In this respect, MPDs resemble localized orbitals.

\subsubsection{Different structures, similar MPDs}
\begin{figure}[htb]
 \begin{center}
   \includegraphics[width=0.75\textwidth]{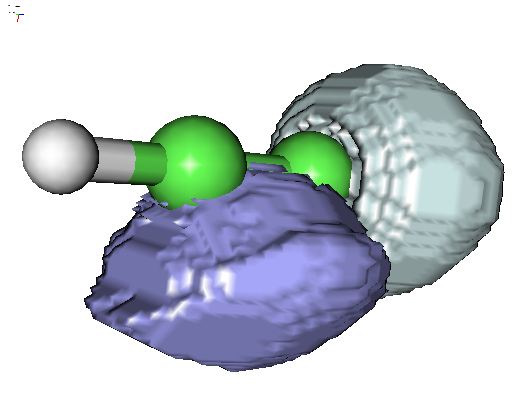} \\
   \includegraphics[width=0.75\textwidth]{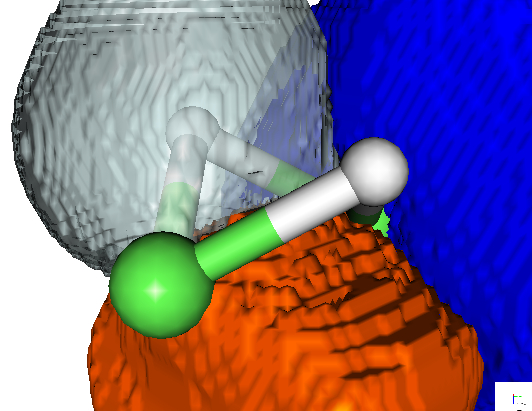} 
 \end{center}
 \caption{MPDs in acetylene \ch{C2H2}, top, and \ch{Si2H2}, bottom. The MPDs containing the H nuclei are silver-colored; one of the triple banana bond MPDs in \ch{C2H2} is shown in purple; the lone pair of Si is shown dark blue while the bent Si-Si bond MPD is shown in red.}
 \label{fig:c2h2-si2h2}
\end{figure}

Fig.~\ref{fig:c2h2-si2h2} shows MPDs for \ch{C2H2} and \ch{Si2H2}.
Some of the MPDs are not shown for clarity; they can be easily obtained by symmetry operations.
We know that the most stable structure of \ch{C2H2} is linear, while for \ch{Si2H2} we have a "butterfly" structure~\cite{LeiKraFre-05}.
With MPDs, however, we get a different perspective.
In both cases, we find three electron pairs between the heavy atoms, and one electron pair pointing out from the other heavy atom.
In \ch{C2H2}, the first three correspond to the three "banana" bonds, and the last to the CH bond.
In \ch{Si2H2}, the first three correspond to the two three-center SiHSi bonds, and one SiSi bond, while the latter corresponds to a lone pair.
It is as if electron pairs like a tetrahedral arrangement, and nuclei arrange to fit into it.
The probabilities are 0.4 for the bent bond regions (CC and SiSi), and 0.5 for the regions corresponding to the other electron pairs.
Apparently there is an extra localization provided by the protons.

\clearpage
\subsubsection{Selecting the relevant region}

\begin{figure}[htb]
 \begin{center}
   \includegraphics[width=0.75\textwidth]{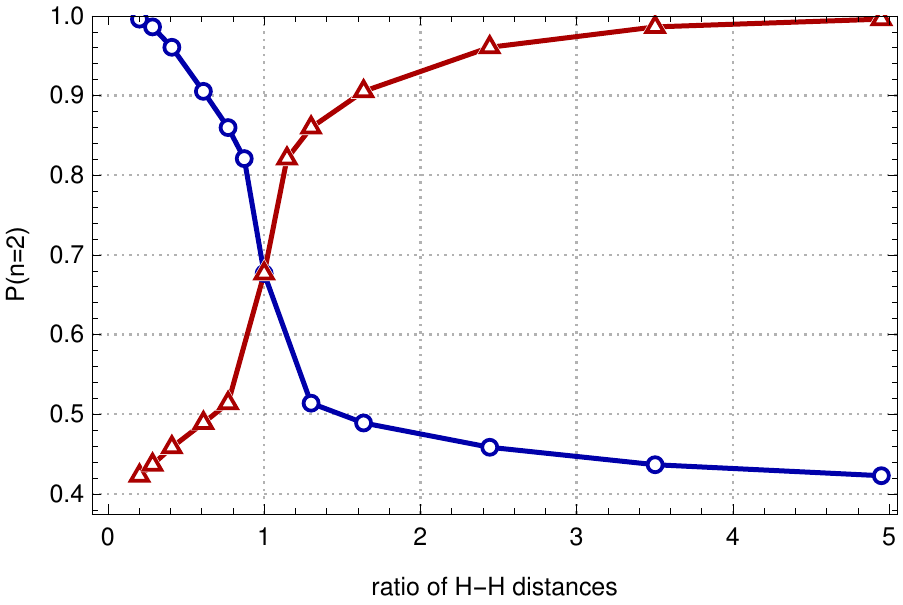} 
 \end{center}
 \caption{Probability to have 2 electrons in the lower (or upper) half-space (blue), or left (or right) half-space (red) as function of the deformation of the rectangular arrangement of the nuclei; on the abscissa the ratio of the sides of the rectangle.}
 \label{fig:p-h4}
\end{figure}

During a chemical reaction MPDs evolve.
Let us consider, for example, the potential energy surface of the following reaction
\begin{equation}
\begin{array}{ccc}
\text{H} & - & \text{H} \\
 \vdots & & \vdots  \\
 \vdots & & \vdots  \\
\text{H} & - & \text{H} 
\end{array}
\; \longrightarrow \;
\begin{array}{ccc}
\text{H} & - & \text{H} \\
| & & | \\
\text{H} & - & \text{H} 
\end{array}
\; \longrightarrow \;
\begin{array}{ccc}
\text{H} & \dots\;\;  \dots & \text{H} \\
| & & | \\
\text{H} & \dots\;\; \dots& \text{H} 
\end{array}
\end{equation}
For a rectangular arrangement of the nuclei, we divide the space symmetrically into a region containing the upper two H nuclei, $\Omega_{\text{up}}$, and one containing the lower two H nuclei, $\Omega_{\text{down}}$. 
We can also divide it into a left and right region ($\Omega_{\text{left}}, \Omega_{\text{right}}$).
The probability to find two electrons for the region where the H nuclei are closer to each other is higher than that for the other division.
For the first structure indicated above on the left, we have
$P(n=2,\Omega_{\text{up}}) = P(n=2,\Omega_{\text{down}}) > P(n=2,\Omega_{\text{left}}) = P(n=2,\Omega_{\text{right}})$
while for the structure shown on the right,
$P(n=2,\Omega_{\text{up}}) = P(n=2,\Omega_{\text{down}}) < P(n=2,\Omega_{\text{left}}) = P(n=2,\Omega_{\text{right}})$
The transition between the two "best" choices occurs at the square arrangement.
The evolution of probabilities is shown in figure~\ref{fig:p-h4}. 
It shows that passing through the square region can be associated with a change of the chemical description.

\subsubsection{The effect of the Pauli principle}

$P(n,\Omega)$ can be significantly larger than that obtained using a binomial distribution,
\begin{equation}
    P_{ind}(n,p)= \frac{N!}{n! (N-n)!} p^n (1-p)^{N-n}
\end{equation}
This distribution is obtained when considering that each of the $N$ electrons would have the probability $p$ to be in $\Omega$.
For some choices of $n$ and $\Omega$, 
\[ P(n,\Omega_n) > \max_p P_{ind}(n,p) \]
i.e., the wave function produces a larger probability than the largest that could be produced by statistically independent particles.
The main reason behind it is the Pauli principle.

\begin{figure}[htb]
 \begin{center}
   \includegraphics[width=0.95\textwidth]{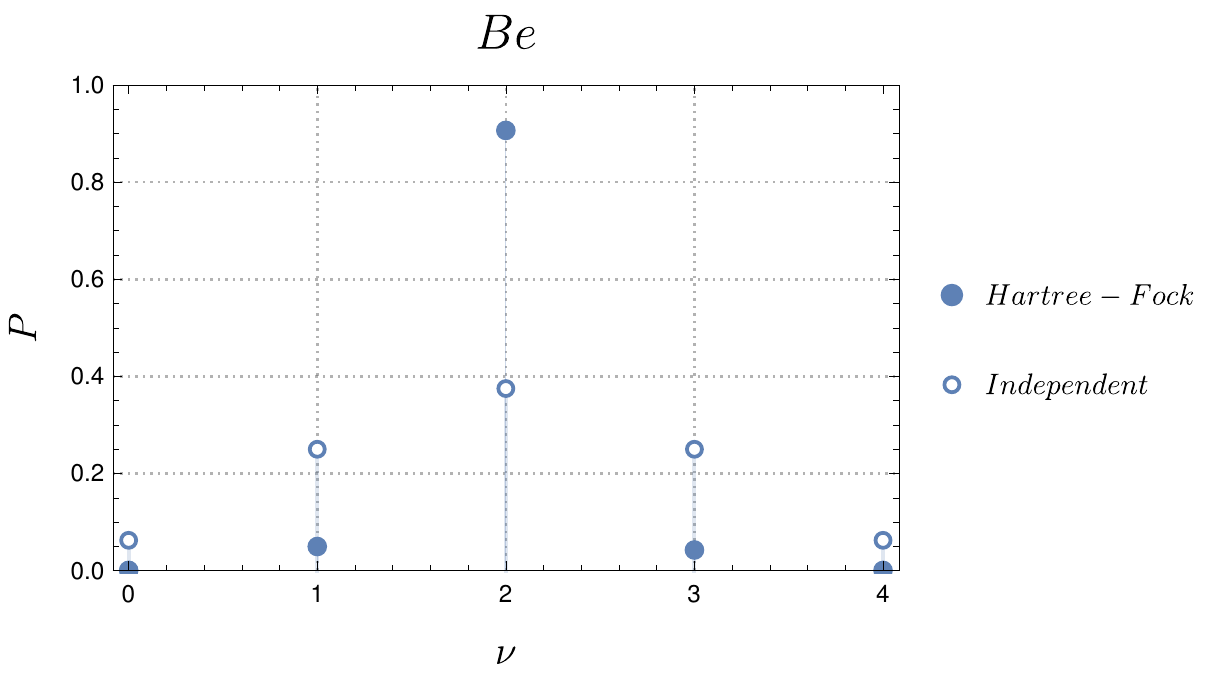}
 \end{center}
 \caption{Probability distribution for the MPD corresponding to the valence shell of Be (full circles) compared to that of independent particles (empty circles).}
 \label{fig:p-be}
\end{figure}

Let us consider as an example the Be atom.
We separate it into two regions, an inner sphere, corresponding to the core, and the rest of the space, corresponding to valence.
All interpretative models give the sphere a radius of $\approx$1~bohr.
Fig.~\ref{fig:p-be} shows the probability distribution obtained when making a core/valence separation with the MPDs.
It is compared with that would be obtained for independent particles, namely that obtained with a binomial distribution (producing the highest
possible outcome for 2 electrons in each of the regions, $p=1/2$).
One clearly sees a higher probability of
 having 2 electrons in each of the shells when using the Hartree-Fock wave function, that satisfies the Pauli principle.

\subsection{Disturbing results}

In addition to the potential of providing "chemical" answers using quantum mechanical calculations, the more detailed character given by the probability distributions raises also some questions.

\subsubsection{Low probabilities}

When we define a spatial domain, $\Omega \in \R^3$, quantum mechanics tells us that electrons can cross its limits. 
Even when the average number of electrons in the region corresponds to our expectation, we know that electrons can get into the domain, or out of the domain.
In most cases, there is a non-negligible probability to find a number of electrons different from the chemically expected one.
Let us recall the procedure used.
When constructing an MPD we consider $P(n=n,\Omega)$ where $n$ corresponds to chemical intuition, and find $\Omega_n$, the region that maximizes $P(n,\Omega)$.
For example, choosing $n=2$ we can find in the methane molecule a region for a CH bond, that gives $P(n=2,\Omega_2)$.
Although this is the best (highest) number we can get for the probability, we find numerically, even for good wave functions, that it is only slightly above 1/2.
This means that quantum mechanical fluctuations induce almost the same probability to have a smaller, or a larger number of electrons in this spatial domain.
The dominating contribution comes from having $n\pm1$ electrons in $\Omega_n$.

In the water molecule, the probability to  have two electrons in the lone pair (or the OH bond) is even smaller than 1/2 (the probability to have a number of electrons <2, or >2 is larger than that of having just 2 electrons in it). 
Nevertheless, the population of the MPD is close to 2, because the probability to have $n<2$, electrons, or $n>2$ are nearly equal.

However, the Slater determinant can be built from orbitals having nodes in the same spatial domain.
When we cut out a spatial region, say, a spherical shell in an atom, it is possible to different orbitals to coexist.
For example, the $3d$ orbitals can penetrate the region mainly occupied by the $4s$ orbitals,
and this can explain increasing fluctuations between the M and N shells in Zn.
Let us look closer at the numbers.

\subsubsection{Strong fluctuations}
Let us construct the MPD of an atomic valence shell, i.e., find $\Omega$ as extending from some radius to infinity, such that $P(n=N_{val},\Omega)$ is maximal, $N_{val}$ being the expected number of electrons in the valence shell.

\begin{figure}[htb]
 \begin{center}
   \includegraphics[width=0.45\textwidth]{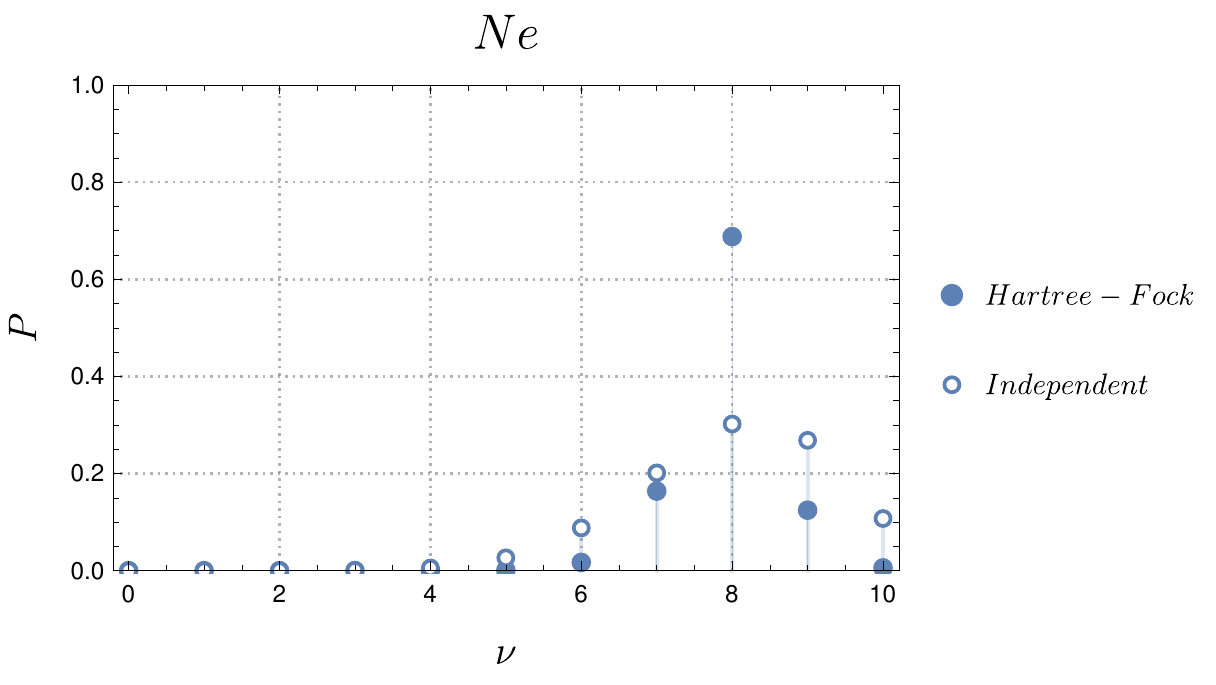}
   \includegraphics[width=0.45\textwidth]{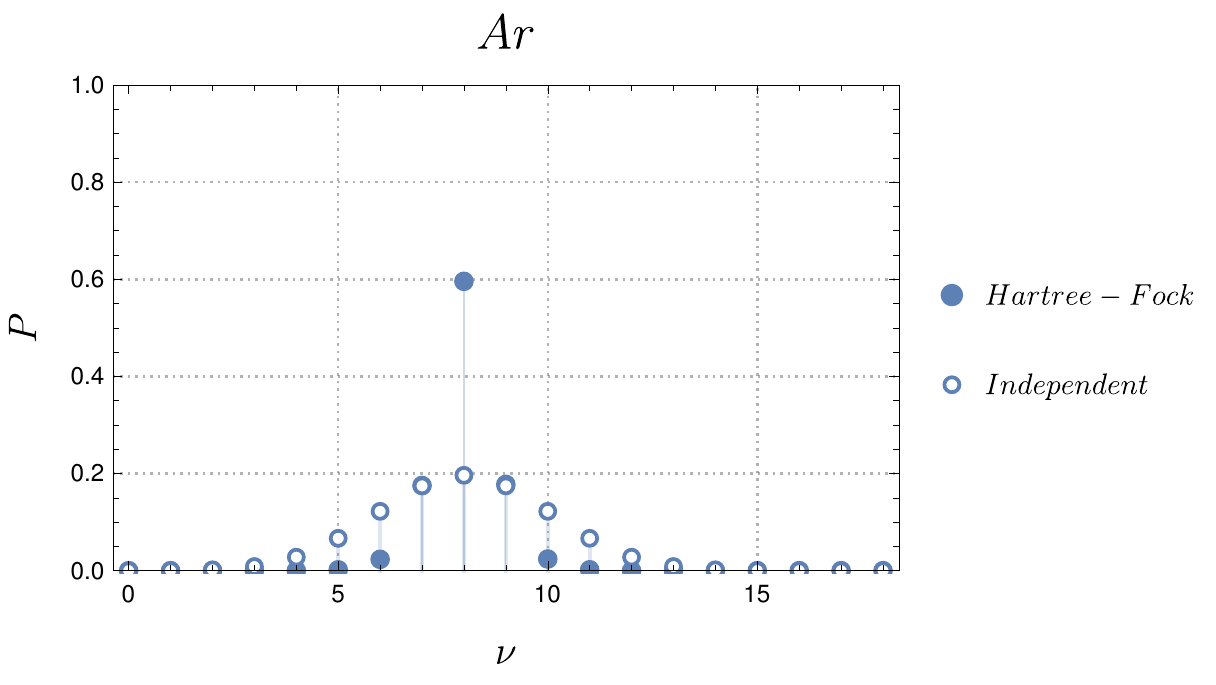} \\
   \includegraphics[width=0.45\textwidth]{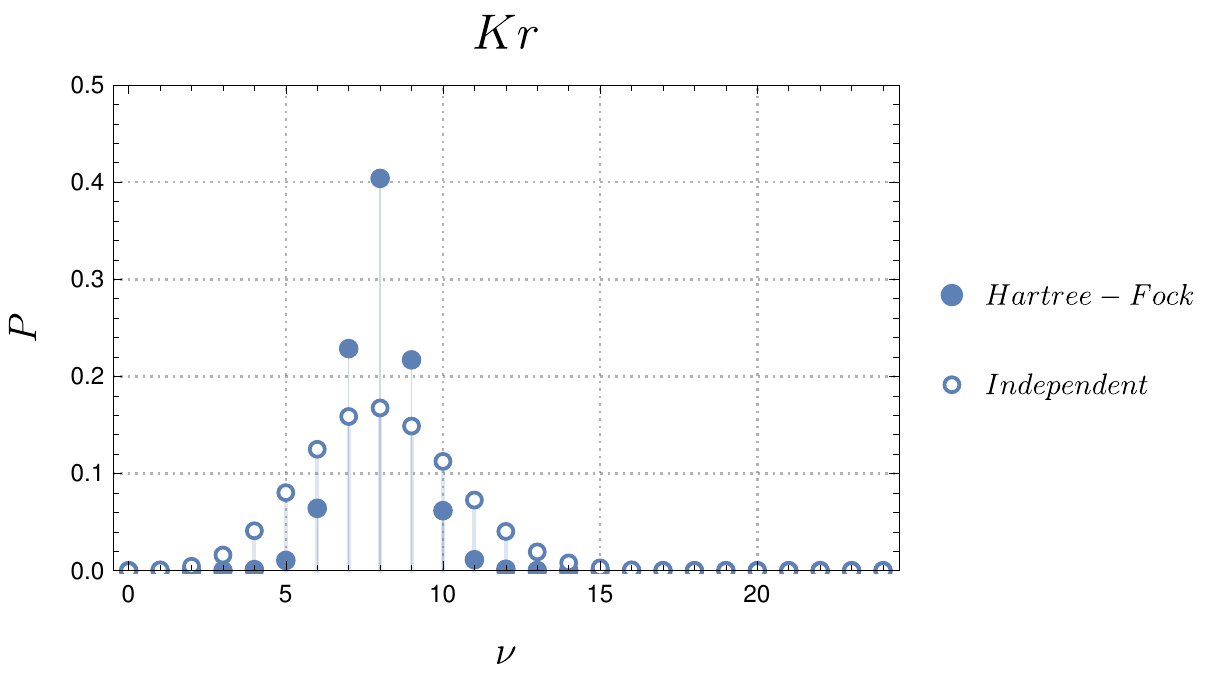} 
   \includegraphics[width=0.45\textwidth]{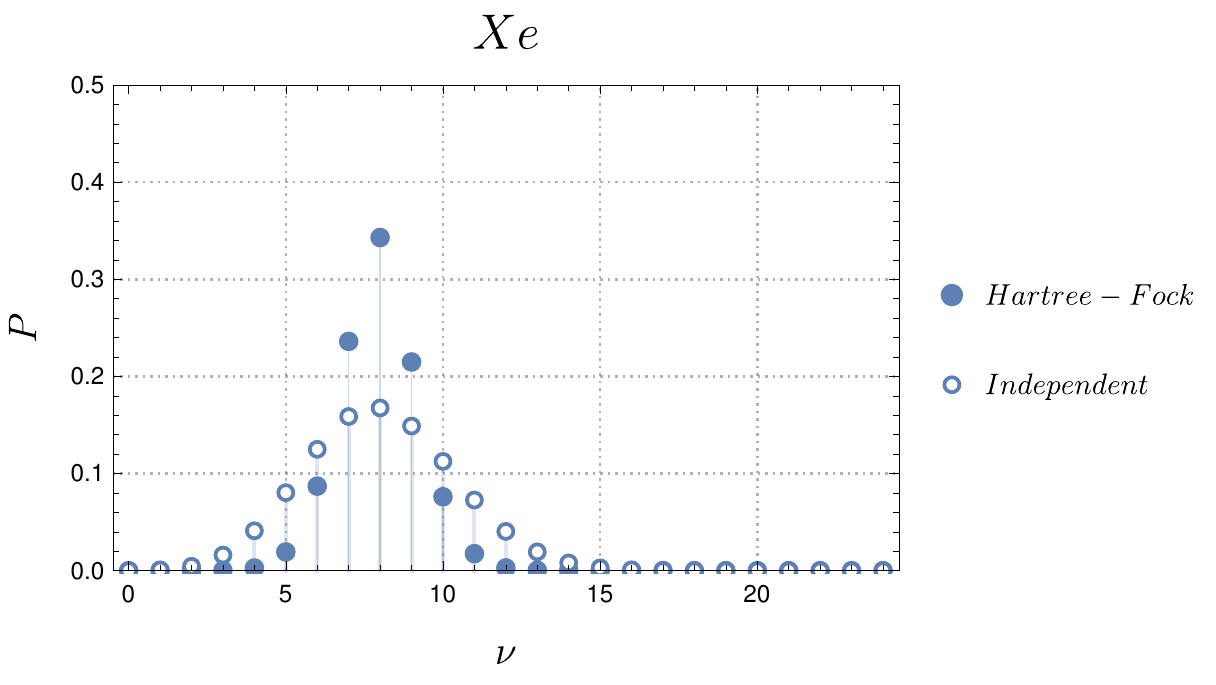} 
 \end{center}
 \caption{Probability distributions for the MPDs corresponding to the valence shells of noble gas atoms (full circles). For comparison, the probability distributions for independent particles obtained for 8 electrons in the valence shell bring able to exchange electrons with the next deeper shell (empty circles).}
 \label{fig:p-noble}
\end{figure}

Let us consider the noble gas atoms, Fig.~\ref{fig:p-noble}.
For Ne and Ar, only the probability of having $N_{val}=8$ electrons in the valence shell is clearly higher than that for independent particles.
This changes for Kr and Xe: we note that finding $N_{val} \pm 1$ electrons in it
is more probable than what one expects for independent particles (exchanging with the deeper shell).
We can attribute the increase in $P(n=8\pm1,\Omega_8)$ to the penetration of the $d$ orbitals of the deeper shell into the valence shell.
The Pauli principle is satisfied not by spatially separating the electrons into regions, but within the same region of space.

\begin{figure}[htb]
 \begin{center}
   \includegraphics[width=0.95\textwidth]{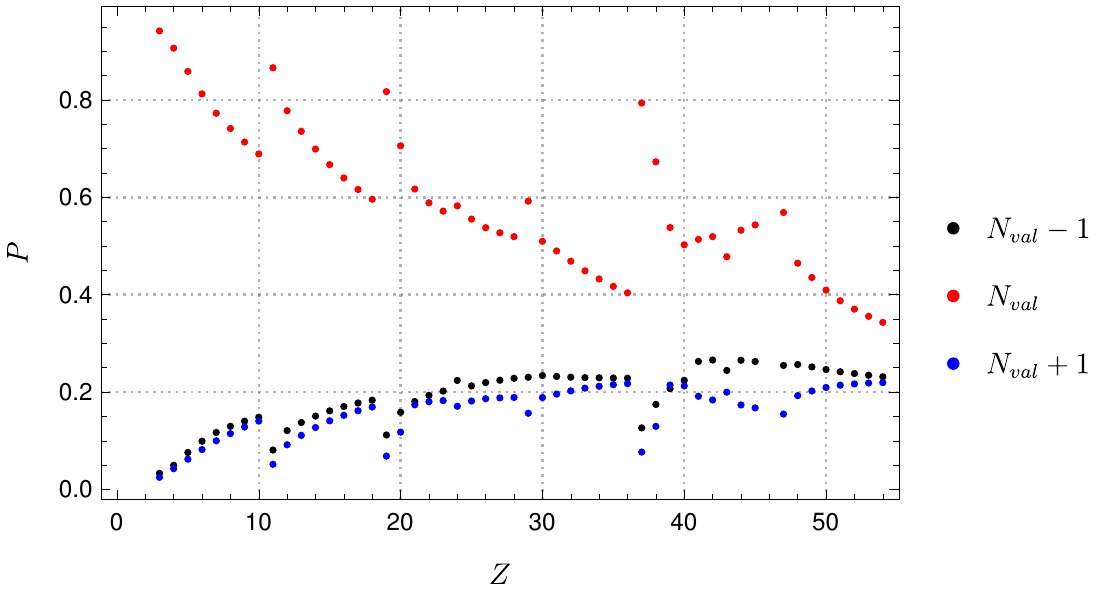}
 \end{center}
 \caption{Probability to have a number of electrons equal to the expected one (red), larger by one (blue), or smaller by one (black) in the MPD corresponding to the valence shell of atoms with nuclear charge $Z \le 54$; H, He, and Pd ($Z=0,1$ or $46$) are not shown, as the results correspond to trivial expectations: the MPDs correspond to the whole space (for $1s^n$), or vanish ($5s^0$).
          }
 \label{fig:ppp}
\end{figure}

Let us analyze the probability of having $N_{val} \pm 1$ electrons in the periodic table (Li-Xe) for Hartree-Fock wave functions~\cite{BunBarBun-93}, cf. Fig.~\ref{fig:ppp}.
One notices a certain symmetry of the distribution: the probabilities of having $N_{val}-1$ or $N_{val}+1$ electrons are, in general, quite close.
If one considers the probability to have not only $N_{val}-1$, or $N_{val}+1$ electrons, but $n < N_{val}$, or $n > N_{val}$ electrons, one obtains in the worst case studied (Xe) an almost equal number for the three probabilities: $P(N > N_{val}) \approx P(N < N_{val}) \approx  P(N_{val})$.
As the MPD is the spatial region yielding the highest possible value for $P(N_{val})$ this casts a shadow of doubt on our image of spatially separated valence shells.

In analogy with valence bond, let us consider the atom formed by a core, $C$, and a valence shell $V$.
In this spirit, one could write:
\[ C^+V^- \leftrightarrow C V \leftrightarrow C^-V^+ \leftrightarrow \dots \]
%{\tt PP: Please add a few clarifying sentences here.}
to indicate that electrons can quit and enter a specific region.
The charges indicated are produced by the separation into shells.
In contrast to valence bond methods, this does not invoke changing orbital occupancies as in valence bond methods.
(Recall that our results are obtained from a Hartree-Fock wave function with a prescribed orbital occupancy.) 
In analogy to valence bond methods it is possible to indicating weights.
Here these are given by the probabilities, e.g., that for $C^+V^-$ is $P(N_{val}+1,\Omega_{\N_{val}})$.
For example, we see in Fig.~\ref{fig:p-noble} that the probability to have 9 electrons in the valence shell of Kr is around 1/4, that we can interpret as a "weight" of $C^+ V^-$. 

Such stronger fluctuations do not occur only in atoms.
For example, the values for the probabilities obtained for the MPDs corresponding to the six electrons of the triple bond in HCCH or N$_2$
are quite comparable to those obtained for the valence shell of Xe.
In HCCH the probability to have six electrons between the two C atoms is around 1/3,
while that to have five (or to have seven) electrons in the same region is
around 1/4.~\cite{SceSav-22}
%\remind{\tt \\ PP: Can you please show data? The overall conclusion of all of
%this is very important and has repercussions on textbook chemistry, not just
%on a single atom.}
In the \ch{N2} molecule, the probability to find 2 electrons in the lone pair is around 0.45,
and in a banana bond only 0.34.
The probability to have only one electron in the banana bond is slightly lower (0.32), and that of having three electrons in it 0.17.

\subsubsection{Choosing the relevant object of study}
\begin{figure}[htb]
 \begin{center}
   \includegraphics[width=0.95\textwidth]{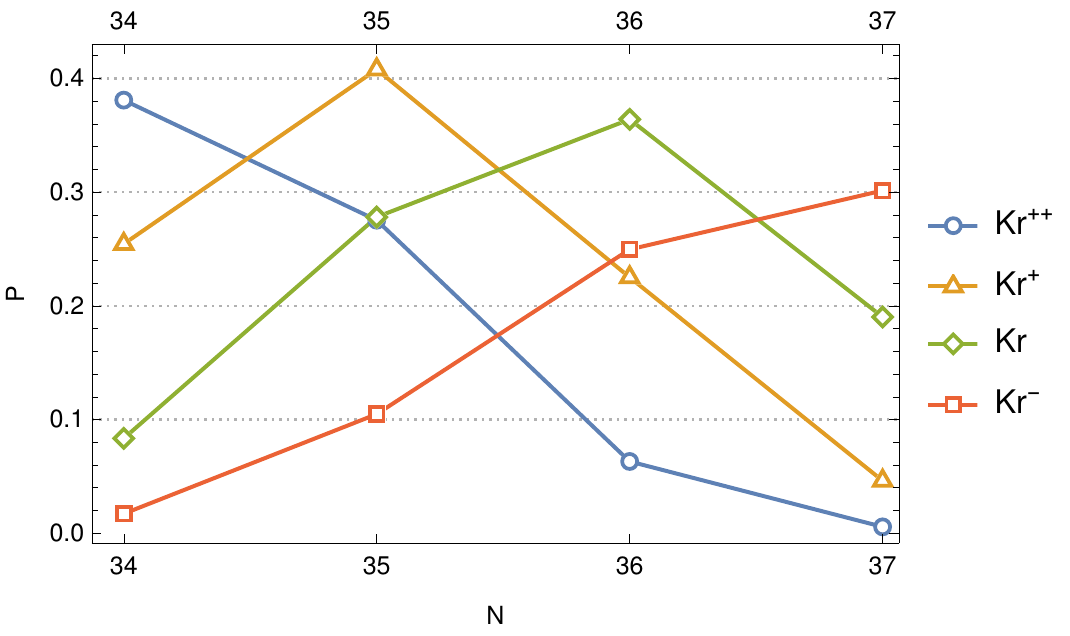}
 \end{center}
 \caption{Probabilities to have $n=34,35,36$, or $37$ electrons in the domain associated to \ch{Kr^{++}} (blue, rhombus), \ch{Kr+} (yellow, triangle), Kr (green rhombus), \ch{Kr-} (orange, square); points are connected by lines to guide the eye.}
 \label{fig:kr-choices}
\end{figure}

Sometimes chemical intuition guides us well in guessing a good number of electrons.
For example, when we are interested in describing an atom, we know its nuclear charge, and it seems natural to choose the same numbers of electrons.
However, would it not be better to sometimes choose an ion?
Let us consider the \ch{KrF2} molecule.
For the Kr atom, we should choose $n=36$, while for \ch{Kr-}, \ch{Kr+}, \ch{Kr^{++}} we should choose $n=37,35$, or $n=34$, respectively.
We take two planes perpendicular to the molecular axis, at equal distance $z$ from the Kr nucleus.~\footnote{For a discussion of the choice of the domains, see Ref.~\cite{SceSav-22}.}
Figure~\ref{fig:kr-choices} shows the probabilities to have $n=34,35,36$ or 37 electrons between the two planes.
We see that there is no clear-cut preference for choosing Kr as a reference: the best (the highest probability) we can get is not better than the one obtained for the separation into \ch{Kr+}, or \ch{Kr^{++}}.
Once we have made the choice, the answers are different.
If we choose the Kr domain, we obtain a probability of 0.36 to describe the region as a Kr atom ($n=36$), and 0.22 as a \ch{Kr+} ion ($n=35$).
If we choose the \ch{Kr+} domain, we obtain a probability of 0.40 to describe the region as a \ch{Kr+} ion, and 0.28 as a Kr atom.

\section{Questions of attitude}

\subsubsection{Three practical questions}
\begin{itemize}
    \item Are MPDs ready for "mass production"? \\
    To obtain MPDs there is a need for new algorithms and programs. Progress is made, but slowly. 
    Furthermore, the existing programs take some time for the optimization of the spatial domain, and this is opposed by all those who think that it is worth having a long quantum mechanical calculation, but not for producing an interpretation using it.
    Evidently, the present authors do not share this opinion.
   \item How much input is needed from the user ? \\
   Some methods (QTAIM, ELF), just let the program work (maybe with a little help). With MPDs, the users have to specify the number of particles they are interested in, an initial guess of the region where the MPD is of interest.
   \item When is an interpretative method that we, theoreticians,  propose successful? \\
   When experimentalists use it. 
   With MPDs we are not yet so far.
\end{itemize}

\subsubsection{Do we need MPDs?}
We could imagine that our computers could give, e.g., by machine learning all the answers to the questions we would like to ask.
Would it be sufficient?
One would like the interpretative methods give tools to let us think independently of the computer.

The next question is whether we should accept the computer help us to think about chemistry.
Maybe a common answer is that given by Prof. C. Pisani (University of Torino) when he criticized ELF: ``With MO theory, you can help yourself using the back of an envelope''. 
Here is a philosophical support for this attitude of independence of external support.
\begin{quote}
 Socrates. At the Egyptian city of Naucratis, there was a famous old god, whose name was Theuth [Toth]; \dots \, he was the inventor of many arts, such as arithmetic and calculation and geometry and astronomy and draughts and dice, but his great discovery was the use of letters. Now in those days the god Thamus [Amun] was the king of the whole country of Egypt \dots. To him came Theuth and showed his inventions, desiring that the other Egyptians might be allowed to have the benefit of them; he enumerated them, and Thamus enquired about their several uses, and praised some of them and censured others, as he approved or disapproved of them. \dots But when they came to letters, This, said Theuth, will make the Egyptians wiser and give them better memories; it is a specific both for the memory and for the wit. Thamus replied: \dots  this discovery of yours will create forgetfulness in the learners' souls, because they will not use their memories; they will trust to the external written characters and not remember of themselves. The specific which you have discovered is an aid not to memory, but to reminiscence, and you give your disciples not truth, but only the semblance of truth; they will be hearers of many things and will have learned nothing; they will appear to be omniscient and will generally know nothing; they will be tiresome company, having the show of wisdom without the reality. \\
 
 Plato, \emph{Phaedrus}\footnote{Plato, Phaedrus 274b, translated by B. Jowett,  \url{http://classics.mit.edu/Plato/phaedrus.html}}
\end{quote}
The present authors are full of admiration for those who are able to use only the back of the envelope. However,
\begin{itemize}
 \item experience accumulated using computers may help developing such methods,
 \item nowadays, we live with Wikipedia in our pocket and it is a good starting point for our thinking;
 the computers can give us ideas we can think about.
\end{itemize}

\section{Conclusion}
The present paper considers the probabilities to have a chosen number of electrons in a spatial domain.
If these domains are optimized in the sense of maximizing the probability, they can be associated to classical chemical concepts.
For example, one would consider two electrons for defining a region of a lone pair, or that of a single bond.
As the probabilities
\begin{enumerate}[label=\roman*]
    \item are not needed to high accuracy, and 
    \item have second order errors when the departure from the optimal domain is of first order,
\end{enumerate}
high accuracy in optimization is not needed.

Sometimes the chemical question is not well set. 
For example, should we define an atomic region, and look at the probability of having a number of electrons different number of electrons in it, or should we start by first defining an ionic region?
The results obtained are not the same.
Furthermore, the highest probability to have a chemically significant electron number in a given spatial domain is often not far away from that of having a different number of electrons in the same region.
In some cases the probability is higher. 
For example, the probability of having 2 electrons in the core and the rest in the valence for the atoms decreases from 0.9 to 0.7 from Li to Ne.
However, one is used to consider a statistical event relevant if the probability is higher than 0.95 ($\approx 2\sigma$ for a normal distribution). 
This was never observed in the systems presented here.
Does this mean that we should give up the chemical concepts associated to a given number of electrons in a spatial region?
The main argument for not doing so is the success of the chemical concepts.
Did we not look at the right quantities? Finally, they seem to be recovered in an average sense, because the distribution of probabilities are often symmetric around the maximum, mean values, {\it i.e.}, populations, are most often used in discussions.
However, we should not forget the quantum nature produces more information, and it may be worth looking into it its implications.

\section*{Acknowledgement}
The authors are grateful to Pascal Pernot (Université Paris-Saclay) for stimulating discussions, and to the editor of the present volume, Paul Popelier (University of Manchester), for improving our manuscript.

\appendix
\section{Mathematical definition of the Maximum Probability Domains}
\label{app:mpd}

Maximum Probability Domains (MPDs) are spatial regions that maximize the probability of having a chosen number of particles, $n$, in them.~\cite{Sav-02}
For discussing the chemical bonding in molecules (see, e.g., ~\cite{GalCarLodCanSav-05}), crystals (see, e.g., \cite{CauSav-11}) the particles considered are the electrons. 
However, there are applications, where the particles have different nature, e.g., for solvation,  the particles considered may be atoms, ions, molecules, \dots.~\cite{AgoCicSavVui-15}

For Slater determinants, in the limiting case that the localization of orbitals is perfect (no overlap between them), the MPDs are identical~\cite{Sav-05} to the spatial domain where these orbitals are localized, or the basins~\cite{SilSav-94} of the electron localization function, ELF~\cite{BecEdg-90}.
The probability to find $n$ electrons in the spatial region $\Omega$ is given by:
\begin{equation}
P(n,\Omega)=\left( \begin{array}{c} N \\ n \end{array} \right)
       \int_\Omega d1 d2 \dots dn \int_{\bar{\Omega}} dn+1 \dots dN 
    \mid \Psi \mid^2
\label{eqn:p}
\end{equation}
Here, $\Psi$ is the wave function of the system and $N$ the number of electrons.
The integration over the region $\Omega$ is performed for the first $n$ electrons. 
The integration is performed over the remaining space, $\bar{\Omega}$, for the other $N-n$ electrons.
The prefactor arises because the electrons are not distinguishable: any other choice of $n$ electrons contributes (by the same amount) to $P(n,\Omega)$.
The MPD is the spatial region that maximizes, for a given $n$, $P(n, \Omega)$,
\begin{equation}
 \Omega_n = \arg\max_\Omega P(\Omega, n)
\label{eqn:mpd} 
\end{equation}
Note that $\Omega_n$ may be a collection of spatially disconnected domains.

Obtaining $P(n,\Omega)$ seems difficult, except for Quantum Monte Carlo calculations where one has only to count how many times $n$ electrons are in $\Omega$.~\cite{SceCafSav-07}
An algorithm for computing $P(n,\Omega)$
using only the overlap between occupied orbitals, can be found in~\cite{CanKerLodSav-04},
and extensions to multi-determinant wave functions exist.~\cite{FraPenBla-07}
One can also work with models, e.g., the Hubbard model.~\cite{AckDeBClaVanPoeVanBul-16}

Maximizing the probability can be done by different algorithms. One can divide space, define a collection of these spatial elements for $\Omega$ and add or eliminate spatial elements to reshape the spatial domain to maximize the probability.~\cite{MafBraCauSav-11}
There are more refined methods, like the level set method.~\cite{CanKerLodSav-04}

\section{Details of the underlying computations}
\label{app:comput}
The atomic calculations were performed using the Hartree-Fock wave functions of ref.~\onlinecite{BunBarBun-93}.
The calculations for \ch{Si2H2} were performed at the Hartree-Fock level with the energy-consistent pseudopotentials (and corresponding basis sets) of the Stuttgart/Cologne group~\cite{BerDolKueStoPre-93}.

For \ch{F2}, \ch{H2O}, \ch{C2H2} and \ch{KrF2} we used the electron
configurations generated for Ref~\onlinecite{SceSav-22}. These were obtained by
sampling wavefunctions generated with the CIPSI algorithm in the valence full
CI space.

In the geometries used for \ch{H4}, the H-H bond lengths (in \AA{}) are obtained as
\begin{eqnarray}
r_1 = 1.276 - 0.135275 \, n \\
r_2 = 1.276 + 0.590525 \, n
\end{eqnarray}
where $n=0$ corresponds to the transition state and $n=4$ corresponds to a
geometry optimized at the CAS(4,4)/cc-pVDZ level.  At each geometry, the wave
function was computed with the cc-pVTZ basis set close to the full
configuration interaction (CI) level ($E - E_{\text{FCI}} < 10^{-4}$ au) using
a wave function made of determinants selected with the CIPSI algorithm. The
energies are given in table~\ref{tab:h4_energies}.

\begin{ruledtabular}
\begin{table}[]
    \centering
    \caption{Geometries and energies for the potential energy curve of \ch{H4} $\longleftrightarrow$ 2\ch{H2}}
    \label{tab:h4_energies}
    \begin{tabular}{lcccc}
       $n$  & $r_1$ (\AA)  & $r_2$ (\AA) & Energy (au) & $E_{\text{FCI}}$ (au) \\
    \hline
       0.00     & 1.2760  & 1.2760  & -2.10783 & -2.10786 \\
       0.25     & 1.2422  & 1.4236  & -2.13344 & -2.13345 \\
       0.50     & 1.2084  & 1.5713  & -2.16824 & -2.16826 \\
       1.00     & 1.1407  & 1.8665  & -2.22134 & -2.22136 \\
       2.00     & 1.0055  & 2.4571  & -2.28642 & -2.28644 \\
       3.00     & 0.8702  & 3.0476  & -2.32775 & -2.32779 \\
       4.00     & 0.7349  & 3.6381  & -2.34459 & -2.34460 \\
    \end{tabular}
\end{table}
\end{ruledtabular}

The CIPSI calculations were made using the Quantum Package
program\cite{GarAppGas-19}, and the electron configurations were sampled
with the the QMC=Chem code\cite{SceCafOse-13}.

\bibliography{biblio-proba}{}
\end{document}